\documentstyle[11pt,epsf,axodraw]{article}

\input{epsf.sty}

\def\fo{\hbox{{1}\kern-.25em\hbox{l}}}

\def\slashchar#1{\setbox0=\hbox{$#1$}           
   \dimen0=\wd0                                 
   \setbox1=\hbox{/} \dimen1=\wd1               
   \ifdim\dimen0>\dimen1                        
      \rlap{\hbox to \dimen0{\hfil/\hfil}}      
      #1                                        
   \else                                        
      \rlap{\hbox to \dimen1{\hfil$#1$\hfil}}   
      /                                         
   \fi}                                         %
\renewcommand{\theequation}{\arabic{section}.\arabic{equation}}
\def\mEt{\mbox{${\hbox{$E$\kern-0.6em\lower-.1ex\hbox{/}}}_T$}\, } 
\def\plusconj{{}}
\def\displ{{}}
\def\Omeg{X}
\def\bfy{{\bf y}}

\def\la{\langle}
\def\xx{x}
\def\ra{\rangle}

\newcommand{\newc}{\newcommand}
\newc{\lcal}{\int {\cal L}dt}
\newc{\gsim}{\lower.7ex\hbox{$\;\stackrel{\textstyle>}{\sim}\;$}}
\newc{\lsim}{\lower.7ex\hbox{$\;\stackrel{\textstyle<}{\sim}\;$}}

\newc{\mw}{M_W}
\newc{\msusy}{m_{W}}
\def\beq{\begin{eqnarray}}
\def\eeq{\end{eqnarray}}
\def\bea{\begin{eqnarray}}
\def\eea{\end{eqnarray}}
\catcode`@=11
\long\def\@caption#1[#2]#3{\par\addcontentsline{\csname
  ext@#1\endcsname}{#1}{\protect\numberline{\csname
  the#1\endcsname}{\ignorespaces #2}}\begingroup
    \small
    \@parboxrestore
    \@makecaption{\csname fnum@#1\endcsname}{\ignorespaces #3}\par
  \endgroup}
\catcode`@=12

\begin{document}
\begin{titlepage}
\begin{flushright}
hep-ph/9907550 \\
FERMILAB-Pub-99/198
\end{flushright}
\vspace{0.3in}

\baselineskip=20pt
\begin{center}
{\Large\bf \hbox{Dimensionless supersymmetry breaking couplings,
flat directions,}
and the origin of intermediate mass scales}\\
\end{center}
\large

\vspace{.15in}
\begin{center}

{\sc Stephen P.~Martin}

\vspace{.1in}
{\it Department of Physics, Northern Illinois University, DeKalb IL 60115
{$\rm and$}\\
}{\it Fermi National Accelerator Laboratory,
P.O. Box 500, Batavia IL 60510 \\} 

\vspace{0.8in}

{\bf Abstract}
\end{center}

The effects of supersymmetry breaking are usually
parameterized by soft couplings of positive mass dimensions. However,
realistic models also predict the existence of suppressed, but
non-vanishing, dimensionless supersymmetry-breaking couplings. These
couplings are technically hard, but do not lead to disastrous quadratic
divergences in scalar masses, and may be crucial for understanding
low-energy physics. In particular, analytic scalar quartic couplings that
break supersymmetry can lead to intermediate scale vacuum expectation
values along nearly-flat directions. I study the one-loop effective
potential for flat directions in the presence of dimensionless
supersymmetry-breaking terms, and discuss the corresponding
renormalization group equations. I discuss two applications: a minimal
model of automatic $R$-parity conservation, and an extension of this
minimal model which provides a solution to the $\mu$ problem and an
invisible axion.  

\end{titlepage}

\baselineskip=17.4pt
\setcounter{footnote}{1}
\setcounter{page}{2}
\setcounter{figure}{0}
\setcounter{table}{0}
\vfill
\eject

\section{Introduction}
\label{sec:intro}
\setcounter{equation}{0}
\setcounter{footnote}{1}

The primary motivation for supersymmetry (for reviews, see
\cite{NHK,primer})
as an extension of the Standard Model is that it can stabilize the hierarchy
associated with the electroweak scale. This implies that supersymmetry is
softly broken \cite{HK,gg}.  The technical version of this requirement is
often taken to mean that quadratic divergences in radiative corrections to
scalar masses must be absent to all orders in perturbation theory. However,
the determination of which couplings are soft then depends on whether the
theory contains any gauge-singlet chiral superfields which can engender
tadpole loop diagrams. This is a rather obscure criterion from the
low-energy point of view, since the minimal supersymmetric standard model
(MSSM) contains no such fields, but reasonable extensions of it often do.
Furthermore, if supersymmetry is spontaneously broken, then the low-energy
theory will generally include couplings that are not soft according to the
technical definition. In realistic models, the hard supersymmetry breaking
couplings are usually expected to be highly suppressed, and that is why they
are traditionally neglected. In this paper, I will explore in detail some
circumstances in which ``hard" supersymmetry breaking couplings nevertheless
are important, and even crucial, for understanding physics at low energies.
 
Let us suppose that the spontaneous breaking of supersymmetry can be
parameterized by the vacuum expectation value (VEV) of an auxiliary $F$-term
for some chiral superfield $X$. The appearance of supersymmetry breaking in
the low-energy theory can then be understood as coming from
non-renormalizable Lagrangian terms which couple $X$ to the other chiral
superfields $\Phi$ and gauge field-strength superfields $W^a_\alpha$ in the
theory. These terms are suppressed by powers of some large mass scale $M$,
which in supergravity-mediated supersymmetry breaking is related to the
Planck mass. The relevant couplings of lowest dimension include,
schematically:\footnote{The subscripts $D$ and $F$ in eq.~(\ref{softorigin})
correspond to integrations $\int d^4\theta$ and $\int d^2\theta$
respectively.  Indices $a,b,c\ldots$ are used for gauge adjoint
representation indices, $\alpha = 1,2$ is a two-component fermion index, and
indices $i,j,k\ldots$ will be used for chiral supermultiplet labels.  The
chiral superfields $\Phi$ have complex scalar and fermion components $\phi$
and $\psi$, and $\lambda^a$ denote gaugino fields.}
\beq
-{\cal L} = 
\left ({1\over M} [X W^{\alpha a} W^a_\alpha]_F
+ {1\over M}[X \Phi^3]_F 
+ {\mu\over M}[X \Phi^2]_F 
\right ) + {\rm c.c.} +
{1\over M^2} [X^* X \Phi^* \Phi]_D .
\label{softorigin}
\eeq
So, with $\langle \Omeg \rangle = \theta\theta F $,
one finds the usual soft terms
\beq
-{\cal L} = \left ( 
{F\over M} \lambda^a \lambda^a
+ {F\over M} \phi^3 
+ {\mu F\over M} \phi^2\right ) 
+ {\rm c.c.} +
{|F|^2 \over M^2} \phi^* \phi ,
\eeq
corresponding 
to gaugino masses, 
cubic scalar couplings, analytic scalar squared masses and
non-analytic squared masses. We can therefore make the identification
$F/M \sim m_W$, indicating that these soft terms are very roughly of order
the
electroweak scale. 
The analytic scalar squared mass is parameterized here as being
proportional
to a corresponding superpotential mass parameter $\mu$. This is not
strictly necessary, but corresponds to the phenomenologically viable
picture of the MSSM in which $\mu$ is itself expected to be of order
$F/M$ or 1 TeV, much less than the high scale $M$. (In a more complete
model,
$\mu$ is likely to be replaced by the VEVs of some other fields so that
it is related to $F/M$ in some way.)

However, there is no good reason why the only terms that couple $X$ to the
other fields should be the ones shown in eq.~(\ref{softorigin}). All
possible renormalizable supersymmetry-breaking terms can arise from
appropriate non-renormalizable supersymmetric couplings involving one or
two powers of
$X$. This is shown in Table 1, in which the interactions are written
schematically along with their possible origin and softness
according
to the technical definition. I list each coupling as either ``soft" (if
it never leads to quadratic divergences in conjunction with renormalizable
supersymmetric couplings), or ``maybe soft" (if no quadratic divergences
can be induced 
provided that
there are no gauge-singlet chiral superfields), or ``hard". The last
column lists the lowest-dimension possible origin for the term. The
subscript $D$ or $F$ indicates whether the original term involves
$\int d^4\theta$ or $\int d^2 \theta$. The
penultimate column then lists the
resulting naive suppression, assuming that the coefficient of the original
term involving $X$, $X^*$ is of order unity. Of course, this last
assumption can easily be violated by symmetries, so that the actual
suppression may be stronger or weaker than the naive estimate shown.
\renewcommand{\arraystretch}{2.0}
\begin{table}[tp]
\caption{Classification of all renormalizable supersymmetry breaking
interactions. 
Chiral scalars and fermions are represented by $\phi$ and $\psi$, and
gauginos by $\lambda$.
The last column indicates the lowest dimension operator
which can give rise to the term through spontaneous supersymmetry breaking
with $\langle X \rangle = \theta\theta F$ and a high scale of
suppression $M$. The resulting naive suppression is shown in the third
column. 
\label{tab:classification}}
\vspace{0.4cm}
\centerline{
\begin{tabular}{|c|c|c|c|}
\hline
Type & Term & Naive Suppression & Origin
\\
\hline\hline
&$\phi \phi^* $ &
$\displ {|  F  |^2\over M^2}\,\sim\, \msusy^2 $ &
$ {1\over M^2}[ \Omeg\Omeg^* \Phi\Phi^* ]_D$
\\  \cline{2-4}
soft &
$\phi^2 \plusconj $ & 
${\mu F\over M} \sim \mu m_W$
& ${\mu\over M} [ \Omeg \Phi^2 ]_F$
\\  \cline{2-4}
& $\phi^3 \plusconj $&
$\displ {  F   \over M} \,\sim\, \msusy $ &
$ {1\over M} [ \Omeg \Phi^3 ] _F$
\\  \cline{2-4}
& $\lambda\lambda \plusconj $&
$\displ {  F   \over M} \,\sim\, \msusy $ &
$ {1\over M} [\Omeg W^\alpha W_\alpha ] _F$
\\  \hline  \hline
&$\phi^2 \phi^* \plusconj $&
$\displ {|  F  |^2 \over M^3}\,\sim\, {\msusy^2 \over M}$ &
$ {1\over M^3}[ \Omeg\Omeg^* \Phi^2 \Phi^* ]_D$
\\  \cline{2-4}
maybe soft &
$\psi\psi \plusconj $&
$\displ {|  F  |^2 \over M^3} \,\sim\,
{\msusy^2 \over M} $ &
$ {1\over M^3}[ \Omeg\Omeg^* D^\alpha \Phi D_\alpha \Phi ]_D$
\\  \cline{2-4}
& $\psi\lambda \plusconj $&
$\displ {|  F  |^2 \over M^3}\,\sim\, {\msusy^2 \over M}$&
$ {1\over M^3}[ \Omeg\Omeg^* D^\alpha \Phi W_\alpha ]_D$
\\  \hline   \hline
&$\phi^4$ &
$\displ {  F   \over M^2}\,\sim\, {\msusy \over M}$
&
$ {1\over M^2}[ \Omeg \Phi^4 ]_F$
\\ \cline{2-4}
& $\phi^3\phi^*$ &
$\displ {|  F  |^2 \over M^4}\,\sim\, {\msusy^2 \over
M^2}$ &
$ {1\over M^4}[ \Omeg\Omeg^* \Phi^3 \Phi^* ]_D$
\\ \cline{2-4}
& $\phi^2\phi^{*2}$ &
$\displ {|  F  |^2 \over M^4}\,\sim\, {\msusy^2 \over
M^2}$ &
$ {1\over M^4}[ \Omeg\Omeg^* \Phi^2 \Phi^{*2} ]_D$
\\ \cline{2-4}
& $\phi \psi\psi$ &
$\displ {|  F  |^2 \over M^4}\,\sim\, {\msusy^2 \over
M^2}$ &
$ {1\over M^4}[ \Omeg\Omeg^*\, \Phi D^\alpha \Phi D_\alpha \Phi ]_D$
\\ \cline{2-4}
hard & $\phi^* \psi\psi$ &
$\displ {|  F  |^2 \over M^4}\,\sim\, {\msusy^2 \over M^2}$ &
$ {1\over M^4}[ \Omeg\Omeg^*\, \Phi^* D^\alpha \Phi D_\alpha \Phi ]_D$
\\ \cline{2-4}
& $\phi \psi\lambda$ &
$\displ {|  F  |^2 \over M^4}\,\sim\, {\msusy^2 \over M^2}$ &
$ {1\over M^4}[ \Omeg\Omeg^*\, \Phi D^\alpha \Phi W_\alpha ]_D$
\\ \cline{2-4}
&
$\phi^* \psi\lambda$ &
$\displ {|  F  |^2 \over M^4}\,\sim\, {\msusy^2 \over M^2}$ &
$ {1\over M^4}[ \Omeg\Omeg^*\, \Phi^* D^\alpha \Phi W_\alpha ]_D$
\\ \cline{2-4}
&
$\phi\lambda\lambda$ &
$\displ {  F    \over M^2}\,\sim\, {\msusy \over M}$ &
$ {1\over M^2}[ \Omeg \Phi W^\alpha W_\alpha ]_F$
\\ \cline{2-4}
&
$\phi^* \lambda\lambda$ &
$\displ {|  F  |^2 \over M^4}\,\sim\, {\msusy^2 \over
M^2}$ &
$ {1\over M^4}[ \Omeg\Omeg^*\, \Phi^* W^\alpha W_\alpha ]_D$
\\ \hline
\end{tabular}
}
\end{table}

The existence and potential importance of the non-analytic cubic couplings
$\phi^2\phi^*$ and chiral fermion mass terms $\psi\psi$ have been
recognized in several papers
e.g.~\cite{HK,Inoue:1982ej,
Jones:1984cu,Hall:1990ac,Borzumati:1999sp,Jack:1999ud}.
These terms
are
closely related to each other, since any chiral fermion mass term
$\psi\psi$ can be absorbed into the superpotential, at the cost of also
redefining the $\phi^2 \phi^*$ couplings if there are superpotential
Yukawa couplings present. In this sense, hard chiral fermion mass terms
are redundant. The chiral fermion gaugino mass mixing terms $\lambda \psi$
can exist if a chiral fermion is in an irreducible representation which 
is also found in the adjoint representation of the
gauge group \cite{Jones:1984cu}. This does not occur in the MSSM, but can
easily happen in extended models. The one-loop renormalization group (RG)
equations for all of these dimensionful supersymmetry breaking couplings
have been recently worked out in ref.~\cite{Jack:1999ud}.

The dimensionless hard supersymmetry breaking couplings include three
distinct types of quartic scalar couplings $\phi^4$, $\phi^3\phi^*$,
$\phi^2
\phi^{*2}$, and various types of scalar-fermion-fermion couplings as
allowed by gauge symmetries. For example, there are $\phi\lambda\lambda$
couplings if a chiral supermultiplet transforms as a representation found
in the symmetric product of the adjoint with itself. There are
non-analytic Yukawa couplings $\phi^* \psi\psi$, as well as non-symmetric
analytic Yukawa couplings $\phi_i\psi_j\psi_k$. Unlike Yukawa couplings
following from the superpotential, the latter need not be symmetric under
interchanges of $i,j$ or of $i,k$. Note that all of the couplings listed
in
Table 1 that have a $D$-term origin are suppressed by\footnote{If $D$-term
VEV(s) plays an important role in
supersymmetry breaking, then $|F|^2/M^2$ can be replaced (or added to) by
$D$ everywhere
in Table 1.} $|F|^2/M^4$, while those that have an $F$-term origin are
only suppressed by $F/M^2$. 

There are at least two reasons why the suppressions of the ``maybe soft"
and ``hard" couplings listed in Table 1 need not render them irrelevant.
First, it may be that our simplest notions of spontaneous supersymmetry
breaking are incorrect, so that the naive suppressions that are
usually implicitly
assumed are not present. For example, it may be that the
high scale $M$ which governs the suppressions is actually far below the
Planck scale. 
Or, several different high scales, some much smaller than others, may
govern the suppression of these terms.
Second, the MSSM and other realistic supersymmetric models
generically have many ``flat directions"
\cite{flatdirections,Dine:1996kz,gkm,stringflats} along
which
the renormalizable
supersymmetric part of the scalar potential vanishes identically. Along
the flat directions, the effects of the usual dimensionful soft terms
become less
relevant at
intermediate and high scales, so that the dimensionless
supersymmetry-breaking couplings become significant despite their
suppression.
  
In this paper, I will concentrate on the dimensionless supersymmetry
breaking couplings of the type $\phi^4$. While these couplings are
technically hard,\footnote{Actually, $\phi^4$
couplings are always technically soft at one-loop order; they only
give rise to quadratic divergences in scalar squared masses at two-loop
order or higher.}
they are soft in the practical sense that they
do arise
generically in models with spontaneous supersymmetry breaking. They are
typically suppressed only by one power of $F/M$, unlike the $\phi^3
\phi^*$ and $\phi^2 \phi^{*2}$ dimensionless terms and the $\phi^2\phi^*$
scalar cubic term.  
Section \ref{sec:RGEs} contains the one-loop RG
running for these couplings, and gives analytic solutions for the case
that gauge couplings dominate, or more generally when the $\phi^4$
term
parameterizes a flat direction of the renormalizable supersymmetric
lagrangian.
These results are useful for relating
high-scale boundary
conditions on the soft terms to the possibility of symmetry breaking at
much lower scales.  
In section \ref{sec:toy}, I will describe how the
supersymmetry-breaking $\phi^4$ couplings are relevant for symmetry
breaking at intermediate scales, including a discussion of the one-loop
effective potential. 
In sections \ref{sec:appsone} and \ref{sec:appstwo}, I will apply these
considerations to
two examples: a minimal model
of $B-L$ breaking to $R$-parity, and an extension of this minimal model
which also incorporates an invisible
axion and a solution to the $\mu$ problem.
Section \ref{sec:conclusions} contains some concluding
remarks.

\section{Running of supersymmetry-breaking $\protect{\phi^4}$ couplings}
\label{sec:RGEs}
\setcounter{equation}{0}
\setcounter{footnote}{1}  

In this section I will discuss the RG equations relevant for running
an $h\phi^4$-type coupling from a high scale (where a boundary condition
on it is to be provided) to an intermediate scale or other scale of
interest. Consider a superpotential of the form
\beq
W = {1\over 24 M} \xx^{ijkl} \Phi_i\Phi_j\Phi_k\Phi_l + {1\over 6 }
Y^{ijk} \Phi_i\Phi_j\Phi_k\> ,
\eeq
with a gauge group with coupling(s) $g_a$.
There are also supersymmetry breaking couplings
\beq
-{\cal L}_{\rm breaking} = \left (
{1\over 24 M} h^{ijkl} \phi_i\phi_j\phi_k\phi_l
+ {1\over 6} a^{ijk} \phi_i\phi_j\phi_k
+ {1\over 2} M_a \lambda^a\lambda^a \right ) + {\rm c.c.} 
+ (m^2)_i^j \phi^{*i} \phi_j ,
\eeq
where $M_a$ is the gaugino mass, $(m^2)_i^j $ is the scalar squared mass,
$a^{ijk}$ is the holomorphic soft (scalar)$^3$ coupling, and $h^{ijkl}$
is the supersymmetry-breaking (scalar)$^4$ coupling.
Note that both the supersymmetric coupling $\xx$ and the
corresponding
supersymmetry-breaking coupling $h$ are defined with a $1/M$ factor
where $M$ is the high mass scale.
Therefore $\xx$ is dimensionless, and $h$ has dimensions of mass
and should be roughly of order $m_W$.
The RG equations can be computed by requiring 
that large logarithms involving the cutoff can be absorbed into
coupling constant and field redefinitions, introducing a renormalization
scale $Q$. 
The $\beta$-function for any running parameter is equal to its
derivative with respect to $t = $ln$(Q/Q_0)$ where $Q_0$ represents
some fixed energy scale. 

As a point of reference, consider first the
well-known one-loop beta functions of
the renormalizable supersymmetric and soft couplings. For the gauge coupling
and
gaugino mass parameter, one has
\beq
16 \pi^2 \beta(g_a^2) = b_a g_a^3; \qquad\qquad
16 \pi^2 \beta(M_a) = 2b_a g_a^2 M_a ,
\label{rgg}
\eeq
with
\beq
b_a \equiv S_a(R) - 3 C(G_a) ,
\eeq
where $S_a(R)$ is the Dynkin index
summed over all chiral supermultiplets and $C(G_a)$ is the Casimir
invariant of the adjoint representation. 
The normalization is such that for $SU(N)$, $C(G_a) = N$ and a fundamental
representation contributes $1/2$ to $S_a(R)$.
The superpotential Yukawa
interactions obey
\beq
16 \pi^2 \beta ( Y^{ijk} ) & = & \left [
{1\over 2} Y^{ipq} Y_{pqn} Y^{njk} 
- 2 g_a^2 C_a(i) Y^{ijk}
\right ]
\> +\> (i\leftrightarrow j)
\> +\> (i\leftrightarrow k) ,
\label{rgy}
\eeq
and the scalar squared masses have beta functions
\beq
16 \pi^2 \beta( (m^2)_i^j ) &=& 
{1\over 2} Y_{ipq} Y^{pqn} (m^2)_n^j
+ {1\over 2} Y^{jpq} Y_{pqn} (m^2)_i^n
+ 2 Y_{ipq} Y^{jpr} (m^2)_r^q
+ a_{ipq} a^{jpq} \nonumber   \\  
&&- 8 \delta_i^j g^2_a C_a(i) |M_a|^2
+ 2 g_a^2 t_i^{aj} {\rm Tr} [ t^a m^2] \> .
\eeq
where the last term explicitly involving the generators of the gauge group
$t^a$
vanishes for non-abelian groups.
Finally, the beta functions for $a^{ijk}$ terms are given by
\beq
16 \pi^2 \beta ( a^{ijk} ) & = & \left [
Y^{ijn}Y_{npq} a^{pqk} + {1\over 2} Y^{ipq} Y_{pqn} a^{njk}
- 2 g_a^2 C_a(i) a^{ijk}
+ 4 g_a^2 C_a(i) M_a Y^{ijk}
\right ]\nonumber   \\  
&&
\> +\> (i\leftrightarrow j)
\> +\> (i\leftrightarrow k)     .      
\eeq
These results employ the standard convention that $Y_{ijk} =
(Y^{ijk})^*$.
If there are several distinct gauge couplings $g_a$, 
then a sum over the index $a$ is implicit where appropriate.

Now with the conventions outlined and illustrated above, consider the
RG running of the $h$ and $x$ couplings. The coupling
$h^{ijkl}$ runs with
renormalization scale according to
\beq
16 \pi^2 \beta ( h^{ijkl} ) &=&  
\Biggl [{1\over 2} h^{ijpq} Y_{pqn}Y^{nkl} 
+ {1\over 2} h^{ikpq} Y_{pqn}Y^{njl} 
+ {1\over 2} h^{ilpq} Y_{pqn}Y^{njk} 
\nonumber
\\
&&+ {1\over 2} Y^{ipq} Y_{pqn} h^{njkl} 
- 2 g_a^2 C_a(i) h^{ijkl}
+ 4 g_a^2 C_a(i)M_a \xx^{ijkl} \Biggr ]
\nonumber
\\ &&+ (i\leftrightarrow j)
+ (i\leftrightarrow k)
+ (i\leftrightarrow l) .
\label{rghgen}
\eeq
The first three terms in eq.~(\ref{rghgen}) are due to vertex
renormalization,
indicative of the fact that supersymmetry has been broken. 
The next two terms have the standard superfield wave-function
renormalization form. 
However,
the last term is somewhat unusual in that it involves a non-renormalizable
coupling $(x^{ijkl})$ contributing to the RG running of a renormalizable
coupling.
\begin{figure}
\centerline{
\begin{picture}(200,160)(-100,-80)
\DashArrowLine(-60,60)(0,0)5
\DashArrowLine(-60,-60)(0,0)5
\DashArrowLine(60,60)(30,30)5
\DashArrowLine(60,-60)(30,-30)5
\ArrowLine(30,30)(0,0)
\ArrowLine(30,-30)(0,0)
\ArrowLine(30,30)(30,0)
\ArrowLine(30,-30)(30,0)
\Photon(30,30)(30,0)43
\Photon(30,-30)(30,0)43
\Vertex(30,0)3
\end{picture}
}
\caption{The terms in the RG equation (\ref{rghgen})
proportional to gaugino mass come from this logarithmically
divergent Feynman diagram.
The dashed lines correspond to scalar fields, the solid lines
to chiral fermions, and the solid line with a wavy line superimposed
corresponds to a gaugino line with a mass insertion.}
\label{fig:rgcon}
\end{figure}
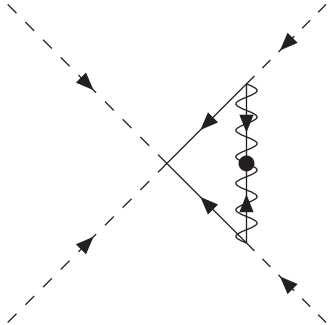
This term comes from the logarithmically divergent Feynman diagram shown
in Figure \ref{fig:rgcon}.
Usually, such contributions are simply ignored, because the
non-renormalizable coupling is understood to be suppressed by the high
scale $M$. However, in the present case it cannot be neglected
because $h$ is also expected to be suppressed. It is easy to see that it
is of the same
order as the other terms, given the estimates 
$h/\xx \sim F/M  \sim M_a \sim m_W$. We will see further evidence for
the necessity of such terms, and the consistency of neglecting other
non-renormalizable effects, in
section \ref{sec:toy}.

The RG running of the corresponding  non-renormalizable superpotential
parameter
$x^{ijkl}$ is just given by superfield wavefunction renormalization:
\beq
16 \pi^2 \beta ( x^{ijkl} ) & = & \left [
{1\over 2} Y^{ipq} Y_{pqn} x^{njkl} 
- 2 g_a^2 C_a(i) x^{ijkl}
\right ]
 + (i\leftrightarrow j)
 + (i\leftrightarrow k)
 + (i\leftrightarrow l) .
\phantom{xxx}
\label{rgxgen}
\eeq

Now, typically the $h\phi^4$ terms are important when investigating
directions
in field space in which the supersymmetric part of the
scalar potential is both $D$-flat and $F$-flat at the renormalizable
level. For this purpose, we can restrict our attention to cases in which
the Yukawa couplings $Y^{ijk}$ do not connect any two fields involved
in the flat direction. In general, $D$-flat directions correspond to
analytic polynomials in the $\phi_i$, which in this case correspond to 
non-vanishing terms
$h^{ijkl}\phi_i\phi_j\phi_k \phi_l$ in the scalar potential.
This
means that when investigating such flat directions, the first three terms
of 
eq.~(\ref{rghgen}) are absent. In many cases, the wavefunction
renormalization factors $Y^{ipq}Y_{pqn}$  will also be proportional to
$\delta^i_n$, so that each
component of $h^{ijkl}$ as well as
$x^{ijkl}$ will evolve separately under RG running.
(An obvious special case of this occurs if the gauge interactions
dominate over the pertinent Yukawa couplings.)
If so, then the ratio of each component of an $h^{ijkl}$ coupling to the
corresponding $\xx^{ijkl}$ coupling will obey a simple RG equation:
\beq
{d\over dt} (h/\xx) = {1\over 4 \pi^2} \sum_a g_a^2 C_i(a) M_a 
+ (i \rightarrow j)
+ (i \rightarrow k)
+ (i \rightarrow l) ,
\label{hoverxrunning}
\eeq
where we omit the indices $i,j,k,l$ in the ratio $h/x$ for simplicity.
The point is that the superfield wavefunction renormalization
factors cancel out of the RG running for the ratios $h/\xx$,
leaving only the terms proportional to gaugino mass. 
This is particularly useful since, as we shall discuss in the next
session, the ratio $h/x$ is most important in deciding whether a
non-trivial minimum can occur along a flat direction.
Equation (\ref{hoverxrunning}) is easily solved in conjunction with
eq.~(\ref{rgg}),
with the result
\beq
h/x = h_0/\xx_0 - 
{(t_0 - t) \over 4 \pi^2}
\sum_a \kappa_a^2 g_{a,0}^2 M_{a,0} C_a(i)
+ (i \rightarrow j)
+ (i \rightarrow k)
+ (i \rightarrow l) ,
\label{ratiorunninggeneral}
\eeq
where 
\beq
\kappa_a(t) = \left [ 1 + 
(t_0 - t)
{g_{a,0}^2 \over 8\pi^2} b_a 
\right ]^{-1/2} ,
\eeq
and $g_{a,0}$, $M_{a,0}$, and $h_0/\xx_0$  are the values of $g_a$,
$M_a$, and $h/x$ at the reference
scale $t_0$.

For example, one might choose to specialize further and assume that the
gauge couplings
and gaugino masses unify to values $g_U$ and $m_{1/2}$ at a unification 
scale $t_0$. Then at lower scales one has the results:
\beq
g_a = \kappa_a g_U\, ;\qquad\qquad
M_a = \kappa_a^2 m_{1/2},
\eeq
If the soft scalar masses have an initial value $m_0^2$ at
the scale $t_U$, then at a lower scale $t$ they are given by the
well-known
result
\beq
m_i^2 = m_0^2 + 2 m_{1/2}^2 \sum_a (1 - \kappa_a^4) C_a(i)/b_a
+ \ldots ,
\eeq
where the ellipses denote negative-definite contributions from
Yukawa couplings. The ratio $h/x$ at the scale
$t$ is just
\beq
h/x = h_0/\xx_0 - 
(t_0 - t) 
{g_{U}^2 \over 4 \pi^2}
m_{1/2} \sum_a \kappa_a^2 C_a(i)
+ (i \rightarrow j)
+ (i \rightarrow k)
+ (i \rightarrow l) .
\label{ratiorunning}
\eeq
The result eq.~(\ref{ratiorunning}) implies that the running of
$h/\xx$
can have either constructive interference, if $h_0/\xx_0$ and
$m_{1/2}$ have opposite signs in our convention, or destructive
interference if $h_0/\xx_0$ and
$m_{1/2}$ have the same sign. (In general, $h/\xx$ and $m_{1/2}$
can be related by a non-trivial complex phase, of course.)
In the case of constructive interference, the dimensionless ratio $h/\xx
m_{1/2}$
can become quite large in the infrared even in the case of a flat direction
made up of fields charged under an abelian symmetry,\footnote{The growth
of $h/\xx m_{1/2}$ can be even more dramatic if the fields participating
in the flat direction are charged under an asymptotically free gauge
interaction; see Appendix A for an MSSM example.}
as we will see in Section 4.
This is the situation that most favors a non-trivial
VEV at an intermediate scale.

\section{Symmetry breaking along flat directions}
\label{sec:toy}
\setcounter{equation}{0}
\setcounter{footnote}{1}  

A toy model which illustrates the essential features we want is as follows.
Consider a pair of chiral superfields which have opposite charges $\pm 1$
under a gauged $U(1)$ symmetry with coupling $g$. 
Assuming that there is a flat direction in the renormalizable,
supersymmetric part of the scalar potential,
the leading contribution to the superpotential can be written as:
\beq
W = 
{x \over 2 M} S^2 \overline S^2 .
\label{wtoy}
\eeq
The coupling $x$ is taken to be real and positive without
loss of generality. 
There is also a $U(1)$ $D$-term contribution,
so that the supersymmetric part of the scalar potential is
\beq
V_{\rm SUSY} = {x^2 \over M^2} |S\overline S|^2
(|S|^2 + |\overline S|^2) + {g^2\over 2}
(|S|^2 - |\overline S|^2)^2 .
\label{vsusytoy}
\eeq
The absence of a superpotential mass term $S\overline S$ can
be ascribed to a discrete symmetry, or
perhaps viewed as a general
result of superstring models which generically do not have
tree-level
masses. 
After spontaneous supersymmetry breaking, there are also
contributions to the scalar potential:
\beq
{V}_{\rm breaking}  = m^2 |S|^2 + \overline m^2 |\overline S|^2
+ \left ({h\over 2M} S^2 \overline S^2 + {\rm c.c.} \right ).
\label{vbreakingtoy}
\eeq
For small field strengths $s$, the soft scalar masses dominate. 
If 
$m^2 + \overline m^2 > 0$, then $s=0$ is a local minimum of the
potential. 
However, the $h$-term contribution to the potential can always be made
negative by
a suitable rephasing of the
fields, without affecting the phases of any of the other terms.
Therefore, the presence of the $h$ couplings always favors the
existence of a non-trivial minimum 
for $S$ and $\overline S$, 
with
spontaneous breakdown of the $U(1)$ symmetry at an intermediate
scale of order $\sqrt{m_W M}$
along the $D$-flat direction.
For the largest field strengths, the superpotential terms
dominate and stabilize the minimum, ensuring that the potential is bounded
from below. 

The $D$-term contribution in eq.~(\ref{vsusytoy}) just forces $S$
and $\overline S$ to have VEVs of very nearly equal
magnitude. Therefore, we can first evaluate the scalar potential in the
approximation that the minimum occurs exactly along the flat direction
$|\langle S \rangle |=|\langle \overline S \rangle |$. (Later in this
section we will discuss the important issue of deviations from
$D$-flatness.)
Parameterizing the flat direction by $s^2 = |S|^2 = |\overline
S|^2$ and $S^2 \overline S^2 = s^4 e^{i\theta}$, the tree-level potential
takes the form
\beq
V_0(s,\theta) = 
(m^2 + \overline m^2) s^2 
+ {1\over 2M}(h e^{i\theta} + h^* e^{-i\theta}) s^4 + 
{2 \xx^2 \over M^2} s^6 .
\label{xpotential}
\eeq
A non-trivial local minimum, if one exists, occurs for $\theta =
\theta_{\rm min}$
with
\beq
h e^{i\theta_{\rm min}} = h^* e^{-i\theta_{\rm min}} = - |h| .
\eeq
Using the estimates 
$h \sim m_W$ and $m^2, \overline m^2 \sim m_W^2$, 
and $\xx$ of order unity, one can check that all of the terms in
eq.~(\ref{xpotential}) are comparable, of order $m_W^3 M$, when $s \sim
\sqrt{m_W M}$. 
The condition
for a local minimum of the tree-level potential is then easily seen to be
\cite{imps,primer,Lazarides:1998iq}:
\beq
|h|^2 - 6 \xx^2 (m^2 + \overline m^2) > 0,
\label{localmin}
\eeq
with 
\beq
s^2 = s_0^2 \equiv {M \over 6 \xx^2} \left [ |h| + \sqrt{|h|^2
- 6 \xx^2(m^2 +  
\overline m^2 )}\right ] \sim  m_W M .
\eeq
Note that 
for smaller
values of $\xx$ the location of the minimum is pushed to
higher scales, but only
the ratio $|h/\xx |$ is important
in deciding whether a non-trivial  local minimum exists. 

In sections \ref{sec:appsone} and \ref{sec:appstwo}, we will discuss
applications in which non-trivial
minima can arise at intermediate scales in the way just described, modulo
certain minor complications. Now, a complete theory of supersymmetry
breaking should predict the values of the supersymmetry breaking
parameters including $m^2$, $\overline m^2$, and $h/\xx$, and therefore in
principle should be able to tell us whether or not a non-trivial
minimum
actually exists. If one has in mind a supergravity-mediated type of model
for supersymmetry breaking, these predictions will take their simplest
form for running parameters near the Planck scale, and will have to be run
down to the intermediate scale. One way to understand this explicitly is
to construct the one-loop effective potential. The result
of doing this is that $h$, $\xx$,
$m^2$, and $\overline m^2$ in the above discussion should be replaced by
running parameters evaluated near the scale of the possible VEV, plus
small calculable one-loop corrections.

It seems worthwhile to see
how this goes explicitly in the present example
by constructing the effective potential along the flat direction.
First, it is important to note that despite the presence of the hard
supersymmetry-breaking coupling $h$, at one loop-order there
is no field-dependent
quadratically-divergent contribution to the scalar potential, since
\beq
{\rm STr} [{\cal M}^2] = 2 (m^2 + \overline m^2) - 2 |M_\lambda|^2 ,
\label{supertracemsquared}
\eeq
even when deviations from $D$-flatness are allowed. 
Here ${\cal M}^2$ are the
eigenvalues of the scalar-field-dependent (mass)$^2$ matrix for the real
scalars,
two-component fermions, and vector bosons in the theory. $M_\lambda$
is the $U(1)$ gaugino mass.
The supertrace
${\rm STr}$ denotes a sum over these modes weighted by $(-1)^s (2s+1)$
where $s$ is the spin of the particle.
Equation (\ref{supertracemsquared})  reflects the fact
that analytic $h\phi^4$ couplings are technically soft at one-loop
order. Field-dependent quadratic divergences do arise at two-loops, but they
are
of order $\Delta V \sim \Lambda^2 |h|^2 \phi^*\phi/M^2$ multiplied by
a two-loop phase-space factor and are therefore safely negligible in the
following one-loop order calculation, as long as the ultraviolet
loop momentum cutoff
$\Lambda$ is not much larger
than the high-scale suppression mass $M$.

There are subtleties involved in evaluating the one-loop effective
potential for arbitrary $S$ and $\overline S$, since the lagrangian
contains non-renormalizable couplings which lead to divergences
in the non-quadratic part of the Kahler potential. We will return to
those issues briefly at the end of this section. However, to a good
approximation,
we can reliably evaluate the effective potential near the flat direction
including only one-loop contributions proportional to $g^2$. Then
the
one-loop effective potential is given in the $\overline {\rm DR}'$ scheme
\cite{DRbarprime}
and in Landau gauge by the usual expression $V = V_0 + V_1$, where 
\beq 
V_1 = {1\over 64 \pi^2} {\rm STr} \left \lbrace
{\cal M}^4 \left [ {\rm ln}({\cal M}^2 / Q^2) - 3/2 \right ]\right \rbrace 
,
\label{onelooppot}\eeq 
with
$Q$ the renormalization group scale. 

The important contributions of order $g^2$ to the one-loop effective
potential 
come from loops involving the massive vector supermultiplet which
arises when $S$ and
$\overline S$ are given VEVs. 
The massive vector supermultiplet
fields consist of one real scalar,
a pair of two-component fermions, and a gauge boson. They have squared 
masses, respectively,
\beq
{\cal M}_0^2 &=& 4 g^2 s^2 + {1\over 2} (m^2 + \overline m^2) -
{1\over 2M} (h e^{i\theta}+ h^* e^{-i\theta}) s^2 - {x^2\over M^2} s^4;\\
{\cal M}_{1/2}^2 &=& 4 g^2 s^2 \pm 2 g s
|M_\lambda e^{i\theta} - {x\over M} s^2|
+ {1\over 2} |M_\lambda|^2 + {x^2 \over 2M^2} s^4; \\
{\cal M}_1^2 &=& 4 g^2 s^2 ,
\eeq
up to terms of order $m_W^3/M$.
Putting these into eq.~(\ref{onelooppot}) and expanding while keeping only
terms which
can contribute proportional to $g^2$ in the scalar potential, one finds:
\beq
V_1(s,\theta) &=& {g^2 \over 16 \pi^2} \Biggl \lbrace 
(4|M_\lambda|^2 - m^2 - \overline m^2) s^2
+ {1\over M} (h e^{i\theta}+ h^* e^{-i\theta}) s^4 + {6x^2 \over M^2}
s^6
\nonumber \\
&&\qquad +{\rm ln}\left ({4 g^2 s^2 \over Q^2}\right )
\Bigl [ (m^2 + \overline m^2 - 8 |M_\lambda|^2 )s^2 
- {10 x^2\over M^2} s^6 \nonumber   \\    &&\qquad
+ {1\over M}
(
4x M_\lambda e^{i\theta} +4x M_\lambda^* e^{-i\theta}
-h e^{i\theta}- h^* e^{-i\theta} 
) s^4
\Bigr ] \Biggr \rbrace .
\label{v1stheta}
\eeq

As a check, we can require that the potential $V_0 +
V_1$
is RG-invariant:
\beq
\left [ 
\beta(x) {\partial\over \partial x} +
\beta(h) {\partial\over \partial h} +
\beta(h^*) {\partial\over \partial h^*} +
\beta(m^2 + \overline m^2) {\partial\over \partial (m^2 +\overline m^2)} 
- s\gamma_s {\partial \over \partial s} 
\right ] V_0
= 
-{\partial V_1\over \partial Q} .\phantom{xx}
\eeq
This equation is solved by
\beq
\beta(x) &=& - {g^2\over 2 \pi^2} x;\\
\beta(h) &=& {g^2\over 2 \pi^2} (2 xM_\lambda - h);
\label{betahconfirm} \\
\beta(m^2 + \overline m^2) &=& -{g^2\over \pi^2} |M_\lambda |^2;\\
\gamma_s &=& - {g^2\over 16 \pi^2} .
\eeq
In particular, eq.~(\ref{betahconfirm}) confirms the RG running for the
$h\phi^4$ couplings found in Section 2.

Now one can minimize the one-loop effective potential $V_0+V_1$ with
respect to $s$ and $\theta$.
The result is that a non-trivial local minimum exists
provided that
\beq
\left [|h|^2 - 6 x^2 (m^2 + \overline m^2)\right ]
\left (1 + {g^2 \over 12 \pi^2} \right ) 
+ {g^2 \over 24 \pi^2} | h + 6 M_{\lambda} x |^2 > 0 ,
\label{oneloopcorrectedcriterion}
\eeq
with all running parameters evaluated self-consistently at 
$Q = 2 g s_{\rm min}$, where $s=s_{\rm min}$ is the resulting minimum.
[Here we drop terms of order $(g^2/16\pi^2)^2$.] So, the region of
parameter space in
which
an intermediate scale VEV is stable is actually increased (and
typically only slightly)
by the one-loop radiative corrections compared to the tree-level
constraints, since
the last correction term
is positive-definite.
The minimum is achieved for
$\theta = \theta_{\rm min}$ satisfying
the same condition $h e^{i\theta_{\rm min}} = h^* e^{-i\theta_{\rm min}} =
- |h|$ as for the
tree-level potential, with $h$
now equal to its running value at $Q = 2 g s_{\rm min}$, and with 
$s_{\rm min}$ explicitly given by
\beq
s^2_{{\rm min}} 
= {M \over 6 \xx^{\prime 2}} \left [ |h'| + \sqrt{|h'|^2
- 6 \xx^{\prime 2}(m^2 + \overline m^2 )'}\right ] ,
\eeq
where
\beq
(m^2 + \overline m^2)' &\equiv& m^2 + \overline m^2 - {g^2 \over 4 \pi^2}
|M_{\lambda}|^2;\\
h' &\equiv& (1 + {g^2 \over 16 \pi^2} )h + {g^2 \over 4 \pi^2} M_\lambda
x;\\
x' &\equiv& (1 + {g^2 \over 24 \pi^2}) x ,
\label{xprimedef}
\eeq
with all quantities on the right side evaluated self-consistently at 
$Q = 2 g s_{\rm min}$.

The one-loop corrections indicated in
eq.~(\ref{oneloopcorrectedcriterion})-(\ref{xprimedef}) are typically
not overwhelming.  
Furthermore, the exact location of $s_{\rm min}$
is not known unless the overall magnitude of $x$ is known, so without
quite detailed model input it 
is not clear at exactly what scale to impose the condition
for a non-trivial minimum. (Specific models of
supersymmetry breaking can predict the ratio
$h/x$ with good accuracy more plausibly than they can predict $h$ or
$x$ separately.) 
Therefore, we are typically justified in 
simply imposing the tree-level conditions for a non-trivial minimum
of the scalar potential, using running parameters evaluated at some
reasonable guess for an
intermediate scale. That is the procedure we will follow in the following
sections.

In the preceding discussion of the one-loop effective potential,
we included only $g^2$ contributions and only field configurations
on the flat direction parameterized by $s$ and $\theta$. It is interesting
to consider the more general case in which these restrictions are not
made. To that end, consider the ${\rm STr} [{\cal M}^4] $ which makes
a contribution to the scalar potential proportional to ${\rm ln}Q^2$.
It is:
\beq
{\rm STr} [{\cal M}^4] &=& 
2 (m^4 + \overline m^4) - 2 |M_\lambda|^4
+ 6 g^2 (m^2 - \overline m^2) (|S|^2 - |\overline S |^2) 
\nonumber  \\   && 
+  2g^2 (m^2 + \overline m^2 -8  |M_\lambda|^2)
(|S|^2 + |\overline S |^2)
+ 8 g^4 (|S|^2 - |\overline S|^2)^2
\nonumber \\  &&
+ {g^2 \over M} ({16 x} M_\lambda - {4h}) S^2 \overline S^2 + {\rm c.c.}
- {20 g^2 x^2\over M^2} (|S|^4 |\overline S|^2 + 
|S|^2 |\overline S|^4)
\nonumber \\ && 
+ {2\over M^2}({ |h|^2 } + {2 x^2 \overline m^2 }) |S|^4 
+ {2\over M^2} ({ |h|^2} + {2 x^2 m^2 }) |\overline S|^4 
\nonumber \\  &&
+ {16\over M^2} [{ |h|^2} + x^2 (m^2 + \overline m^2)] |S|^2 |\overline
S|^2 
+ {20 h x^2\over M^3} (|S|^2 + |\overline S|^2) S^2 \overline S^2+
{\rm c.c.}                                                                 
\nonumber \\ && 
- {4 g^2 x^2\over M^2} (|S|^2 - |\overline S|^2)
(|S|^4 - |\overline S|^4)
+ {16 x^4\over M^4} ( 3 |S|^4 |\overline S|^4 +  |S|^6 |\overline S|^2
+ |S|^2 |\overline S|^6 ) .\phantom{xxxxx}
\label{strmfour}
\eeq
The first few terms, up to 
$(|S|^4 |\overline S|^2 + |S|^2 |\overline  S|^4)$,
have the same form as terms already present in the tree-level scalar
potential. However, the remaining terms do not.
In particular,
the last terms involving 
$(|S|^2 - |\overline S|^2)
(|S|^4 - |\overline S|^4)$
and $( 3 |S|^4 |\overline S|^4 +  |S|^6 |\overline
S|^2
+ |S|^2 |\overline S|^6 )$ do not involve any supersymmetry breaking
couplings and yet appear to have no counterpart in the tree-level     
supersymmetric scalar potential. 

The superpotential is not renormalized. Therefore, to understand the
appearance of these
terms with a ${\rm ln}Q^2$ coefficient, one must include
non-quadratic Kahler potential terms which are renormalized and
whose couplings run logarithmically with RG scale.
The Kahler potential can be expanded in powers
of $1/M$ according to:
\beq
K = S^* e^{2gX} S + \overline S^* e^{-2gX} \overline S 
+ {k_1\over 4M^2}  S^{*2} e^{4gX} S^2 
+ {k_2\over M^2}  S^{*} \overline S^* S \overline S 
+ {k_3\over 4M^2}  \overline S^{*2} e^{-4gX} \overline S^2 +
\ldots\phantom{xx}
\eeq
where $X$ is the vector superfield for the $U(1)$ gauge supermultiplet
and $k_{1,2,3}$ are dimensionless couplings.
There result corrections to the scalar potential:
\beq
\Delta V_F &=&- {x^2 \over M^2} \left [
 (k_1 + k_3) |S|^4 |\overline S|^4 + k_2 |S|^2 |\overline S|^2
(|S|^2 +  |\overline S|^2)^2 \right ] + \ldots
\label{deltavf}
\\
\Delta V_D &=&{ g^2\over 2 M^2}
(|S|^2 - |\overline S|^2)(k_1 |S|^4 - k_3 |\overline S|^4)
+ \ldots \> .
\label{deltavd}
\eeq
The couplings $k_1$, $k_2$ and $k_3$ run with RG scale
according to
\beq
\beta(k_1) = \beta(k_3) = {1\over 2} \beta(k_2) = -{x^2 \over 4 \pi^2},
\eeq
which, together with eqs.~(\ref{deltavf}) and (\ref{deltavd}),
explains the presence of the
last terms in eq.~(\ref{strmfour}).
Note that far from the flat direction,
the non-renormalizable $D$-term contributions in eq.~(\ref{deltavd})
become comparable to all of the other terms in the scalar potential
for intermediate-scale field values.

Fortunately, one can see that at the minimum near the flat direction with
$|\langle S \rangle| = |\langle \overline S \rangle| = s \sim \sqrt{m_W
M}$, all of the ``extra" terms in eq.~(\ref{strmfour}) which were not
reflected in eq.~(\ref{v1stheta}) are suppressed by at least one 
additional factor
of
order $m_W/M$. In checking this, it is useful to note that the
expectation value for
the deviation from $D$-flatness is given by
\beq
|\langle S\rangle |^2 - |\langle \overline  S\rangle |^2 = 
{1\over 2g^2 }(\overline m^2 - m^2) 
+ {1\over 2M^2} (k_3 - k_1) s^4 ,
\label{dtermvev}
\eeq
which is suppressed with respect to $s^2$ by at least $m_W/M$.

We have shown that the existence of an intermediate scale
vacuum expectation value is determined by a one-loop effective
potential which suffers only small corrections with respect to the
tree-level value as long as renormalized couplings near the scale
of the putative VEV are used. In particular, the conditions necessary
for the existence of the intermediate scale VEV do not depend (to the
lowest non-trivial order in $m_W/M$) on non-quadratic Kahler potential
parameters
over which we have little control. On the other hand, eq.~(\ref{dtermvev})
shows that the deviation from $D$-flatness, while appropriately
suppressed,
{\it does} depend on higher-order Kahler potential parameters which
violate
the approximate $S \leftrightarrow \overline S$ symmetry;
both terms in eq.~(\ref{dtermvev}) are of order $m_W^2$.
These $D$-term contributions are important since all scalars $\varphi_i$
in the low-energy theory obtain contributions to their squared masses
equal to
\beq
\Delta m_{\varphi_i}^2 = q_i g^2 (|\langle S \rangle |^2 - |\langle
\overline S \rangle |^2 ) ,
\eeq
where $q_i$ is the $U(1)$ charge of $\varphi_i$. This effect has been
studied in numerous papers including 
\cite{Drees,Hagelin:1990ta,Kawamura:1994yv,Cheng,Kolda}, but
the
presence of
the Kahler-potential-dependent terms in eq.~(\ref{dtermvev}) does not seem
to have been emphasized before. The presence of these terms makes it
very difficult to predict the magnitude or even the sign of the $D$-terms
in the low energy theory, since the Kahler potential terms are generally
not
constrained by symmetries. On the other hand, it also means that the
existence of $D$-terms in the low-energy theory should be quite generic,
and a useful phenomenological prediction of the general scenario espoused 
in this paper.

\section{Minimal model of automatic $R$-parity from gauged $B-L$}
\label{sec:appsone}
\setcounter{equation}{0}
\setcounter{footnote}{1}  

One example of a model in which an $h\phi^4$ term is used to generate
an intermediate scale was proposed in \cite{imps}. The motivation
for this model is as follows.
The MSSM is usually defined to respect $R$-parity (or equivalently,
matter parity) which prevents rapid proton decay and provides for 
a stable neutralino LSP cold dark matter candidate \cite{darkmatter}.
However, in the MSSM the imposition of $R$-parity is somewhat {\it ad
hoc}, since failure to impose it would not result in any internal
inconsistency. The conservation of matter parity
(or equivalently, $R$-parity)
in the MSSM can be explained in terms of a deeper principle by starting with
a continuous gauged $B-L$ symmetry,
where $B$ is baryon number and $L$ is the total lepton number.
Since matter parity is defined to be
\beq
P_M = (-1)^{3(B-L)},
\eeq
a gauged $U(1)_{B-L}$ will forbid all $R$-parity violating operators
\cite{rpbl,ssc}.
There is no massless gauge boson that couples
to $B-L$, so this gauge symmetry must be spontaneously broken.
However, provided that all VEVs or other order parameters in the theory
carry
$3(B-L)$
charges which are {\it even} integers, the $Z_2$ matter parity subgroup
must survive as an unbroken and exact remnant of the original gauged
$B-L$ symmetry \cite{ssc,Kuchimanchi:1995vk,Huitu:1997iy,Aulakh:1999cd}.

In a minimal extension of the MSSM
that realizes this idea, one embeds $U(1)_Y$ into $U(1)_{R} \times
U(1)_{B-L}$, introducing a single new gauge multiplet.
To provide for the breaking 
\beq
U(1)_R \times U(1)_{B-L} \rightarrow U(1)_Y \times Z_2^{{\rm
matter}\>{\rm parity}},
\eeq 
one also introduces
 two chiral superfields $S$, $\overline S$ with $B-L$
charges 
$-2$ and $+2$ respectively. These fields play precisely the same role
as their namesakes in section 3.
In addition to breaking $B-L$ to $R$-parity, an intermediate-scale
VEV for $S$ 
provides a see-saw mechanism for neutrino masses 
by means of superpotential couplings 
\beq
W = {1\over 2} \bfy_S^{ij} S \overline \nu_i \overline \nu_j + 
\bfy_\nu^{ij} \overline \nu_i H_u L_j .
\label{seeandbeseen}
\eeq
Here $\overline \nu_i$ ($i=1,2,3$) are three gauge-singlet neutrinos;
the ordinary Standard Model neutrinos $\nu_i$ are contained in
the $SU(2)_L$-doublet fields $L_i$. 
(The $\overline \nu_i$ must exist anyway in order to provide
for
$U(1)_{B-L}$ anomaly cancellation.)
The resulting $6\times 6$ see-saw neutrino mass matrix in the
$(\nu_i, \overline \nu_i)$ basis is
\beq
-{\cal L} = {1\over 2} \pmatrix{ \nu_i &\overline \nu_i}
\pmatrix{ {\bf 0} & v_u \bfy^T_\nu\cr
v_u \bfy_\nu & \langle S \rangle \bfy_S }
\pmatrix{ \nu_j \cr \overline \nu_j} ,
\eeq
where $v_u = \langle H_u^0 \rangle = 174$ GeV $\sin\beta$.
In order to avoid a disastrously large $U(1)_{B-L}$ $D$-term,
the scalar component of $\overline S $
must get a VEV which is very nearly equal to $\langle S \rangle$.
Many prominent features of this model, including the mass spectrum of the 
$S,\overline S$ supermultiplets, have already been studied
in some
detail in ref.~\cite{imps}, Here, I will briefly review the results found
there, and extend the analysis by examining the solutions to the RG
equations for the $h$-couplings to explore which parts of parameter space
can actually lead to symmetry breaking with $\langle S \rangle \approx
\langle \overline S \rangle$ at an intermediate scale.

The superpotential, supersymmetric scalar potential, and
supersymmetry-breaking terms are all just given
by the same expressions eqs.~(\ref{wtoy}), (\ref{vsusytoy}) and
(\ref{vbreakingtoy})
as
in
section \ref{sec:toy} (with $g\rightarrow g_X$ given below).
In fact, the important features of this model are all identical to
the ones already discussed there, except that the $B-L$ gauge charges
are normalized differently, and there is a non-trivial kinetic mixing
due to the presence of two $U(1)$ gauge symmetries. In the $U(1)_Y$,
$U(1)_{B-L}$ basis, $S$ and $\overline S$ have charges $(0,\pm 2)$.
Another useful basis is the choice $U(1)_R$,
$U(1)_{B-L}$, under which $S$ and $\overline S$ have
charges $(1,-2)$ and $(-1,2)$. However, neither of these bases avoids
kinetic mixing in this model. 
In general, the couplings of the $U(1)$ gauge bosons to a matter field
$\phi_i$ with $U(1)_R$ and $U(1)_{B-L}$ charges $R_i$ and $(B-L)_i$
are given by the covariant derivative
\beq
D_\mu \phi_i &=& (\partial_\mu + i \overline g_i^R A_\mu^R +
i \overline g_i^{B-L} A_\mu^{B-L}) \phi_i,\\
\overline g_i^R &=& g_R R_i + g_{B-L,R} \sqrt{3/8} (B-L)_i,\\
\overline g_i^R &=& g_{B-L} \sqrt{3/8} (B-L)_i + g_{R,B-L} R_i.
\eeq
where $g_{B-L}$, $g_R$, $g_{B-L,R}$ and $g_{R,B-L}$ are
distinct gauge couplings.
In terms of these couplings, the weak hypercharge coupling $g_Y$ and
the coupling
$g_X$ which appears in the $D$-term scalar
potential
$
V_D = {1\over 2} g_X^2 (|S|^2 - |\overline S|^2)^2
$
are given by
\beq
g_X^2 &=& (g_R - \sqrt{3/2} g_{B-L,R})^2 +
(\sqrt{3/2} g_{B-L} - g_{R,B-L})^2 ,
\\
g_Y &=& \sqrt{5/2}(g_R g_{B-L} - g_{B-L,R}g_{R,B-L})/g_X .
\eeq
If one starts at the unification
scale with $SO(10)$-like boundary conditions 
\beq
g_3 = g_2 = g_R =
g_{B-L} = g_U;\qquad\qquad g_{B-L,R} =g_{R,B-L} = 0.
\label{gbcs}
\eeq 
near $Q_0  \sim 3 \times 10^{16}$ GeV, then the successful low-energy
prediction for $\sin^2\theta_W$ is maintained after the symmetry
breaking 
In that case one can most easily run RG equations by going to the
one-loop multiplicatively
renormalized basis, as shown in ref.~\cite{imps}.

For the boundary conditions eq.~(\ref{gbcs}), the analytic one-loop 
results for quantities
of interest are as follows.
The running gauge couplings at any scale $t$ are given by 
\beq
g_R = g_{B-L} = g_U (\kappa_+ + \kappa_-)/2;
\label{gators}\\
g_{R,B-L} = g_{B-L,R} = g_U (\kappa_- - \kappa_+)/2,
\label{huskies}
\eeq
where
\beq
\kappa_{\pm} = \left [ 1 + {g_U^2 \over 8 \pi^2} (9\pm \sqrt{6})
(t_0 - t) \right ]^{-1/2} 
\eeq
are monotonically decreasing as $t$ decreases into the infrared.
[The first equality in each of eqs.~(\ref{gators})
and (\ref{huskies}) is due to an amusing but otherwise inessential 
coincidence in the
matter content of this model.] 
Now let us assume for simplicity that near the gauge coupling
unification scale $t_0$, the supersymmetry breaking masses obey
supergravity
inspired boundary conditions, with a common scalar $m_0^2$, a common
gaugino mass $m_{1/2}$, and a specified $h_0/\xx_0$. 
The running gaugino mass matrix in the $R,(B-L)$ basis is given by
\beq
M_{ab} = {1\over 2} m_{1/2} \pmatrix{
\kappa^2_+ + \kappa^2_- &
\kappa^2_+ - \kappa^2_- \cr
\kappa^2_+ - \kappa^2_- &
\kappa^2_+ + \kappa^2_- }.
\eeq
If the Yukawa couplings in eq.~(\ref{seeandbeseen})
are small enough to be neglected in the RG running, then
the squared masses
for the scalar components of $S$, $\overline S$
run according to
\beq
m_S^2 = m_{\overline S}^2 = m_0^2 + m_{1/2}^2 \left [
\left ( 33+ 13 \sqrt{6} \over 150 \right )(1 - \kappa_+^4)
+ \left ( 33- 13 \sqrt{6} \over 150 \right )(1 - \kappa_-^4)
\right ] .
\label{ranms}
\eeq
The coupling $\xx$ runs according to
\beq
\xx = \xx_0 \,
\kappa_+^{-(66+26\sqrt{6})/75}
\kappa_-^{-(66-26\sqrt{6})/75} ,
\eeq
and, more importantly, the ratio $h/\xx$ runs according to
\beq
h/\xx = h_0/\xx_0 -
(t_0 - t)
{g_U^2 \over 4 \pi^2} m_{1/2}
\left [ 
(5 + 2 \sqrt{6}) \kappa^2_+ +
(5 - 2 \sqrt{6}) \kappa^2_- \right ] .
\label{ranh}
\eeq

\begin{figure}[t]
\centering
\epsfxsize=4.1in
\hspace*{0in}
\epsffile{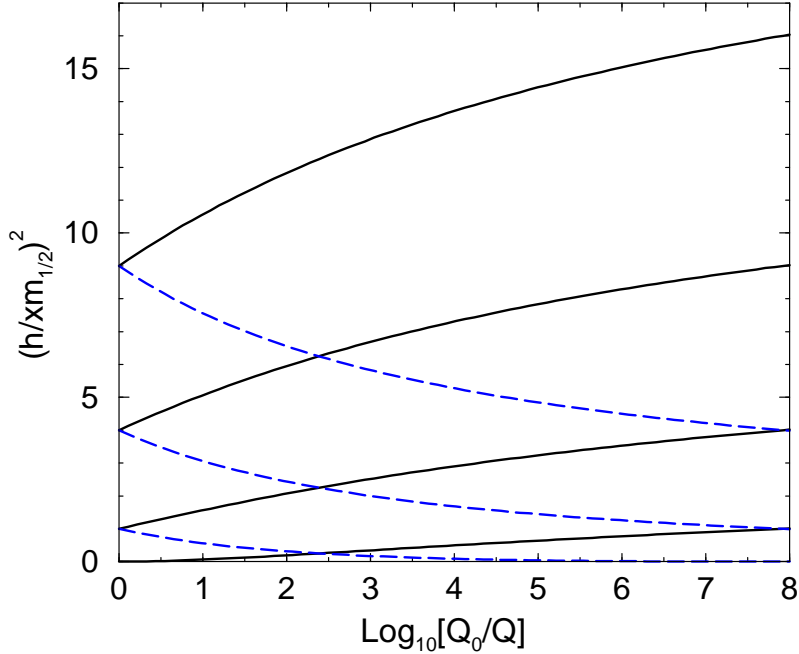}
\caption{RG running of $h/\xx m_{1/2}$ for the minimal model of
automatic $R$-parity conservation described in the text.
Initial values 
at the high scale, corresponding to the left-hand side of the graph,
are $h_0/\xx_0 m_{1/2} = 0,  -1, -2,  -3$ (solid lines) and
$1,2,3$ (dashed lines).
}
\label{rprunning}
\end{figure}

In order to decide whether a non-trivial intermediate scale minimum
$\langle S \rangle 
\approx \langle \overline S \rangle $ will occur, one should compare the
running values of $|h/\xx|^2$ and $m_S^2 + m_{\overline S}^2$
at a scale $Q \sim \langle S \rangle \sim \sqrt{m_W M_P}$, as described in
section \ref{sec:toy}. 
The running of $|h/\xx |^2$ as a function
of scale is shown
in Figure \ref{rprunning}, with a factor of $m_{1/2}^2$ removed to
make it dimensionless. For illustrative purposes, I have chosen initial
values $h_0/\xx_0 m_{1/2} = 0,  \pm 1, \pm 2,  \pm 3$ at the unification
scale, with $g_U^2/4\pi = 0.04$. The figure illustrates that for negative
values of $h_0/\xx_0
m_{1/2}$, there is constructive interference that allows $|h/\xx |^2$
to grow in the infrared (to the right of the figure). The rate of running
gradually slows down as the $U(1)$ gauge couplings diminish at lower
scales. Conversely, positive initial values for $h_0/\xx_0
m_{1/2}$ lead to a destructive interference in the magnitude
of $|h/\xx |^2$ at lower scales. 

\begin{figure}[t]
\centering
\epsfxsize=4.1in
\hspace*{0in}
\epsffile{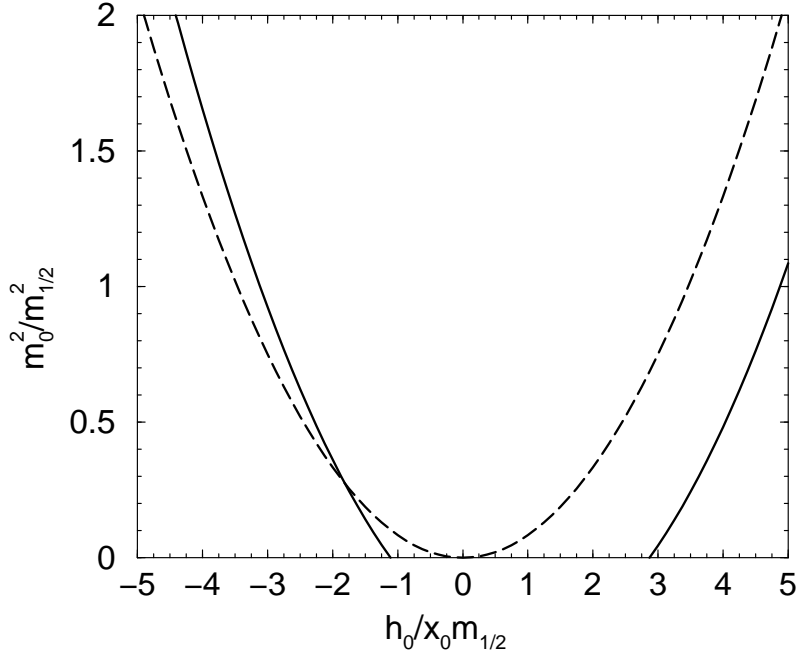}
\caption{The parameter space in the $h_0/\xx_0 m_{1/2}$ vs.
$m_0^2/m_{1/2}^2$ plane for which a non-trivial local minimum occurs along
the $S, \overline S$ flat direction in the model described in the text,
neglecting Yukawa couplings. A local minimum exists for all values of
$m_0^2/m_{1/2}^2$ below the lines shown. The solid line corresponds to a
hierarchy of the high scale to the intermediate scale of $Q_0/Q = 10^6$,
while the dashed line corresponds to the case of no renormalization
($Q_0/Q = 1$).}
\label{testvacgraph}
\end{figure}

Of course, the scalar masses $m_S^2$ and $m_{\overline S}^2$
are also running because of the $U(1)$ gauge couplings. This effect
is quite significant because of the large ($\pm 2$) values of
$B-L$ charges of the
the $S,\overline S$ fields, and it opposes the 
symmetry breaking. 
Therefore, there is a competition between the magnitude of running for
$m_S^2 + m_{\overline S}^2$ and $|h/x|^2$  to decide whether an
intermediate scale VEV can occur. As in section 3, this requires
$|h/x|^2 > 6 (m_S^2 + m_{\overline S}^2)$ at an intermediate scale. 
For a numerical example, one can use $g_U^2/4\pi =
0.04$ and $Q_0/Q = 10^6$ or
$t_0 - t = 13.8$ in eqs.~(\ref{ranms}) and (\ref{ranh}) to get
$m_S^2 = m_{\overline S}^2 \approx m_0^2 + 0.33 m_{1/2}^2$ and 
$h/\xx \approx h_0/\xx_0 - 0.88 m_{1/2}$.
In Figure \ref{testvacgraph}, I show the regions
of parameter space in the $h_0/\xx_0 m_{1/2}$ vs.
$m_0^2/m_{1/2}^2$ plane for which a non-trivial local minimum
occurs along the $S, \overline S$
flat direction in this model. There is a local minimum for all
values of $m_0^2/m_{1/2}^2$ below the lines indicated. The solid
line corresponds to a ratio of the high scale to the intermediate
scale of $10^6$, while the dashed line corresponds to the case of 
no renormalization of parameters (if the intermediate scale were
pushed up close to the high scale, which can occur in the limit
that $\xx$ is very small).
As Fig.~\ref{testvacgraph} shows, the parameter space in which
intermediate-scale symmetry breaking occurs is actually diminished by 
RG running for $h_0/\xx_0 m_{1/2}\gsim -2$, and is increased
for $h_0/\xx_0 m_{1/2}\lsim -2$.
If one assumes a boundary condition $m_{1/2}^2 = 3 m_0^2$ as can
happen in certain dilaton-dominated scenarios of supersymmetry breaking,
then a non-trivial minimum at an intermediate scale requires
$h_0/\xx_0 m_{1/2} \lsim -1.9$ or $\gsim 3.7$ in this picture.

The discussion above is somewhat conservative, in that we have not
included possible negative RG contributions to $m_S^2$ and $m_{\overline
S}^2$.
One might imagine that one could increase the region of parameter space in
which symmetry breaking occurs by including the effects of the Yukawa
couplings $\bfy_\nu$ and $\bfy_S$ in eq.~(\ref{seeandbeseen}). The
latter coupling will certainly act to decrease $m_S^2$ in the infrared,
which clearly favors a VEV for $S$. However, the impact of these
corrections is somewhat limited. 
For example, assuming one neutrino Yukawa coupling $y_S$ and its
corresponding scalar cubic term $a_S$ dominate, the relevant beta
functions are:
\beq
16 \pi^2 \beta(m_{S}^2) &=& 
|y_S|^2 (m_S^2  +  2 m^2_{\overline \nu}) 
+ |a_S|^2 - 8 \sum_a g_a^2 M_a^2 C_a(S)
\\  
16 \pi^2 \beta(m_{\overline \nu}^2) &=& 
2 |y_S|^2 (m_S^2  +  2 m^2_{\overline \nu}) 
+ 2 |a_S|^2 
+\ldots 
- 8 \sum_a g_a^2 M_a^2 C_a(\overline \nu)\phantom{xx}
\label{mountelbert}
\eeq
where the ellipses in eq.~(\ref{mountelbert}) refer to additional
positive-definite contributions from Yukawa couplings. 
A comparison of the terms in these two equations reveals that
without additional interactions, it is
quite difficult to arrange for $m_S^2$ to be driven negative
without first driving ${m}_{\overline \nu}^2$
negative, since $C_a(S) = 4 C_a(\overline \nu)$. Once this occurs, $m_S^2$
will receive large positive
contributions from the 
$2 |y_S|^2 m_{\overline \nu}^2$ term in its beta function. 
One possible alternative is to allow $S$ and $\overline S$ to have
superpotential Yukawa couplings to some heavy particles which also have
gauge interactions with respect to some strongly-coupled gauge
group. The soft squared masses of these new heavy scalars will be
large and positive, and through Yukawa couplings can
allow $m_{S}^2$ and/or $m_{\overline S}^2$ to be driven
negative or at least smaller in the infrared, thus increasing the
parameter space in which intermediate-scale symmetry breaking occurs.
In any case, the contributions of the dimensionless supersymmetry-breaking
terms are likely to be non-negligible, and always favor symmetry breaking.

\section{Model with automatic $R$-parity conservation tied to
a solution to the $\mu$ problem}
\label{sec:appstwo}
\setcounter{equation}{0}
\setcounter{footnote}{1}  

In this section I consider an extension of the
model in the previous section which also incorporates a solution to the
$\mu$-problem together with an invisible axion of the type proposed in
ref.~\cite{kimnilles}.
In
addition to the $U(1)_{B-L}$
gauge supermultiplet and $S,\overline S$ fields, introduce
an additional pair of neutral
chiral supermultiplets $N$ and $\overline N$. These fields are charged
under a global anomalous Peccei-Quinn (PQ) symmetry \cite{PQ} as shown in
Table 2.
When the fields $\langle N \rangle $ and $\langle \overline N \rangle$
get VEVs, they will spontaneously break this PQ symmetry,
which is also explicitly broken by the QCD anomaly.
\begin{table}[b]
\caption{Charges of Higgs fields in the model described in
section \ref{sec:appstwo}. The $B-L$ and $Y$ symmetries are local,
and the PQ symmetry is a global symmetry with a QCD anomaly.
\label{tab:charges}}
\vspace{0.4cm}\centerline{
\begin{tabular}{|c||c|c|c|c|c|c|}
\hline
{}&$S$ & $\overline S$ & $N$ & $\overline N$ & $H_u$ & $H_d$
\\
\hline\hline
$B-L$ & $2$ & $-2$ & $0$ & $0$ & $0$ & $0$\\
\hline
$Y$ & $0$ & $0$ & $0$ & $0$ & $1/2$ & $-1/2$\\
\hline
PQ & $1$ & $-1$ & $-1$ & $1$ & $1$ & $1$\\
\hline
\end{tabular}
}
\vspace{0.2cm}
\phantom{x}
\end{table}
As before, I will assume that superpotential mass terms are absent.
The leading terms in the superpotential consistent with the
symmetries in Table 2 can then be written as:
\beq
W =  {\xx_1 \over 2 M} S^2 \overline S^2
-{\xx_2 \over  M} S \overline S N \overline N
+{\xx_3 \over 2 M} N^2 \overline N^2
+ {\xx_\mu \over M} N^2 H_u H_d .
\label{ww}
\eeq
The last term in eq.~(\ref{ww}) becomes the $\mu$ term of the MSSM
when $N$ obtains its VEV. Then $\langle N \rangle$ should be of order
the intermediate scale $\sqrt{m_W M}$ for two distinct reasons; first
in order to allow the effective contribution to $\mu = x_\mu \langle N
\rangle^2/M$ to be of order
$m_W$, and second to allow the PQ scale to fall within the window 
permitted by direct searches and astrophysical constraints
\cite{axionwindow}.

Because we are looking for a minimum of the potential
with $\la S \ra ,\la \overline S \ra , \la N\ra ,\la \overline N \ra \gg 
\la H_u \ra, \la H_d \ra$, the last superpotential term in 
eq.~(\ref{ww}) is only a small perturbation on the dynamics that
fixes the VEVs of $S, \overline S, N,$ and $ \overline N$. The
relevant
supersymmetric
part of the scalar potential is therefore
\beq
V_{\rm SUSY} &=& 
{1\over M^2} (|S|^2 + |\overline S|^2)|\xx_1 S \overline S - 
\xx_2 N \overline N |^2 
+
{1\over M^2} (|N|^2 + |\overline N|^2)|\xx_3 N \overline N - 
\xx_2 S \overline S |^2 
\nonumber \\   &&
+ {g_X^2\over 2} (|S|^2 - |\overline S|^2)^2,
\label{NSVsusy}
\eeq
and the supersymmetry-breaking terms are
\beq
{V}_{\rm breaking}  &=& 
\left (
{h_1\over 2M} S^2 \overline S^2 
-{h_2\over M} S \overline S N \overline N 
+{h_3\over 2M} N^2 \overline N^2 
\right )
+ {\rm c.c.} 
\nonumber   \\  &&
+ m_S^2 |S|^2 + m_{\overline S}^2 |\overline S|^2
+m_N^2 |N|^2 + m_{\overline N}^2 |\overline N|^2 .
\label{NSVbreaking}
\eeq

The parameter space of this model is multidimensional and complicated,
so I will be content to explore it quantitatively in some special cases.
In order to simplify the analysis, I will 
assume that $\xx_2/\xx_1$ and $\xx_3/\xx_1$ are both real. Then 
each of $\xx_1$, $\xx_2$, and $\xx_3$ can be made real and
positive by a field rephasing, without loss of generality. I will also
assume that 
\beq
h_1/\xx_1 =
h_2/\xx_2 =
h_3/\xx_3 \equiv (h/\xx)_0
\eeq
at the unification scale $Q_0$, and that all scalars have a common
squared mass
equal to $m_0^2$ at that same scale. RG equations are used to run the
parameters
$h_1/\xx_1 m_{1/2}$,
$h_2/\xx_2 m_{1/2}$,
 $m_S^2/m_{1/2}^2$, and
$m_{\overline S}^2/m_{1/2}^2$ 
 to the intermediate scale $Q$  to evaluate the scalar potential.
The ratios $h_3/x_3 m_{1/2}$, $m_N^2/m_{1/2}^2$, and $m_{\overline
N}^2/m_{1/2}^2$ involve gauge singlets and do not run.
Then minima of the potential can be identified by a
straightforward procedure. This involves numerically
solving quartic and quadratic equations to find all
possible candidate extrema, and then requiring that all of the eigenvalues
of the Hessian matrix at the minimum are positive.
The candidate minima occur near the $D$-flat direction
$|\langle S \rangle| = |\langle \overline S \rangle | = s$.
Furthermore, my assumption 
$m_N^2 = m_{\overline N}^2$ implies that $|\langle N \rangle| = |\langle
\overline N \rangle | = n$, because of an approximate symmetry $N
\leftrightarrow \overline N$.

I have checked that for generic choices of $x_1/x_3$ and $x_2/x_3$,
there are stable local minima for sufficiently large values of
$(h/x)_0/m_{1/2}$ compared to $m_0^2/m_{1/2}^2$, just as for the model
in the previous section. However, in the present case both $s$ and $n$ are
non-zero. It
is instructive to first consider the
special case that $Q_0/Q=1$ so that there is no renormalization of
parameters
from the high scale. Then, at least for many values of $x_1/x_3$ and
$x_2/x_3$,  a stable local minimum exists provided that
\beq
(h/x)^2_0 > 
12 m^2_0 
,
\label{norencon}
\eeq 
with $n^2/s^2 = (x_1 + x_2)/(x_3 + x_2)$. 
For the more realistic case
including renormalization effects,
I have chosen to present results for
the special
case $x_1/x_3 = 4$ and $x_2/x_3 = 3$, and for 
case $x_1/x_3 = 1/4$ and $x_2/x_3 = 1/3$,
 in Fig.
\ref{testvacnsgraph}. 
(These ratios are taken to be specified at the intermediate scale.)
For both models, the area below the dashed
curve
is the region of parameter space in which a local minimum exists
when $Q_0/Q=1$, as in eq.~({\ref{norencon}). The area below the
dash-dot line is the corresponding region for $Q_0/Q = 10^6$ when
$(x_1/x_3,x_2/x_3) = (4,3)$. Likewise, the area below the
solid line is the region allowing intermediate scale VEVs
for $Q_0/Q = 10^6$ and $(x_1/x_3,x_2/x_3) = (1/4,1/3)$.
I have checked that similar results obtain for many other choices
of $x_1/x_3$ and $x_2/x_3$.
\begin{figure}[t]
\centering
\epsfxsize=4.1in
\hspace*{0in}
\epsffile{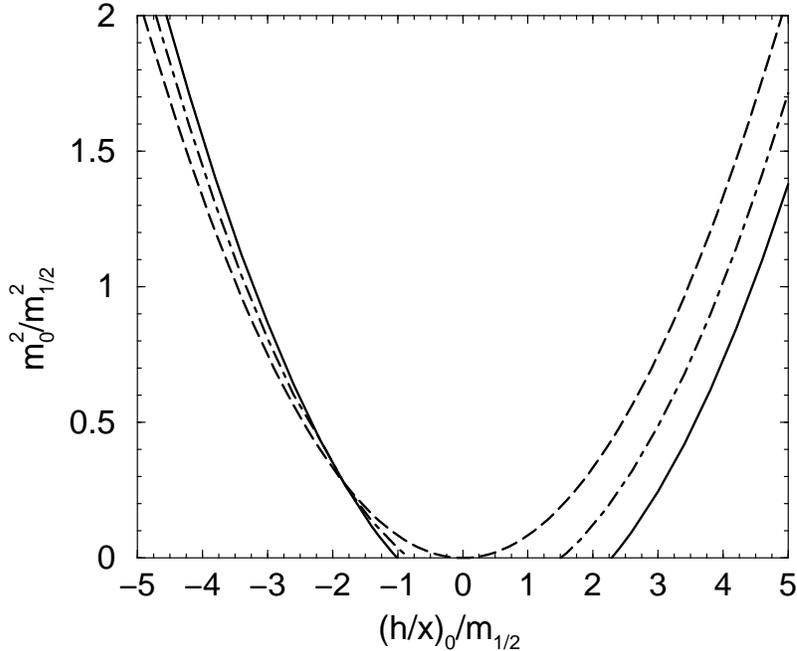}
\caption{The parameter space in the $h_0/\xx_0 m_{1/2}$ vs.
$m_0^2/m_{1/2}^2$ plane for which a local minimum
involving non-zero $\langle S \rangle = \langle \overline S \rangle$
and $\langle N \rangle = \langle \overline N \rangle$ occurs,
for the model and RG boundary conditions described in section
\ref{sec:appstwo}.
At least one non-trivial local minimum exists for all
values of
$m_0^2/m_{1/2}^2$ below the lines shown. 
For a hierarchy of the high scale to the intermediate scale of $Q_0/Q =
10^6$, 
the dash-dot line corresponds to the choice $x_1/x_3 = 4$ and $ x_2/x_3 =
3$ and the solid line corresponds to $x_1/x_3 = 1/4$ and $x_2/x_3 =
1/3$. In the case of no renormalization $(Q_0/Q = 1)$, both models
have non-trivial local minima for all $m_0^2/m_{1/2}^2$ below the dashed
line.}
\label{testvacnsgraph}
\end{figure}

As Fig.~\ref{testvacnsgraph} shows, the critical line for
$(h/x)_0/m_{1/2} < 0$ is surprisingly insensitive to renormalization
group running. This is actually something of an accident, reflecting the
fact that the scalar squared masses and $(h_i/x_i)^2$ run
at comparable rates. For the opposite sign of $(h/x)_0/m_{1/2}$,
there is somewhat more sensitivity.
Also, I find that for
values of $m_0^2$ far below the critical lines shown, there are sometimes
several distinct non-trivial local minima separated by ``ridges" in field
space. Note that the critical lines in this model are between the two
lines shown in Figure~\ref{testvacgraph}. This can be understood as follows:
along the flat direction $(n,s)$ chosen by the model in this section, the
minimum is determined by effective
$h$ and $x$ couplings which are, roughly speaking, ``averages" of some 
couplings that are renormalized by $B-L$ interactions just as in section
4,
and other couplings that are not renormalized because they are gauge
singlets.

However, it is important to realize that the situation depicted in
Fig.~\ref{testvacnsgraph}
is not inevitable, since the relationship between $h_1$, $h_2$, and
$h_3$ can be relaxed. To see this,
note that if the special (one-loop RG-invariant) relationship $
\xx_1 \xx_3 = \xx_2^2$ were to hold, then there would be exactly
flat directions\footnote{Of course, these flat directions will be lifted by
higher-order
terms in the superpotential, but those terms are suppressed by an
additional
factor of $1/M^2$.}
of the supersymmetric part of the  
potential,
eq.~(\ref{NSVsusy}), with $|S|^2 = |\overline S|^2$ and
$N \overline N/S \overline S = \xx_1/\xx_2 = x_2/x_3$.
Now, with the boundary conditions on the $h_i/x_i$ discussed above,
it happens that at one loop order the supersymmetry-breaking quartic part
of
the scalar potential is also flat at one-loop order. However, if one
imagines relaxing this condition and allowing generic values of
$h_i/x_i$, then 
the negative supersymmetry-breaking
$h$-terms will dominate the scalar potential for field values
slightly above the intermediate scale for some
neighborhood of
parameter space near $\xx_1 \xx_3 = \xx_2^2$.
This makes it clear that for even for rather small, but generic,
values of ($h_1$,
$h_2$, $h_3$), there must always be at least some finite region of 
($x_1$, $x_2$, $x_3$) parameter space near $\xx_1 \xx_3 = \xx_2^2$
in which the symmetry breaking does take place. I have checked that
this region of parameter space can be quite large.

The model outlined above has the nice feature that it relates the PQ
intermediate scale governed by $\langle N \rangle$ to the neutrino see-saw
intermediate mass scale governed by $\langle S \rangle$.
A similar model was proposed in
ref.~\cite{Murayama:1992dj}, but in that paper the mechanism for
intermediate-scale symmetry breaking was supposed to be negative RG
corrections to scalar squared masses. In the model described here,
$B-L$ is gauged in order to provide for automatic $R$-parity conservation,
and this makes it extremely difficult for negative scalar squared masses
$m_S^2$ and $m_{\overline S}^2$ to arise. In any case, I have argued that
the $h \phi^4$ coupling
probably provides a significant part of the effect.
If the physical neutrino masses are very small, it may be
required that there is a mild hierarchy 
$
\langle N \rangle \lsim 10^{12}$ GeV $\ll \langle S \rangle 
$ 
in order to accommodate the allowed axion window. This can be accomplished
by choosing the  couplings $x_1, x_2, x_3$ appropriately. 
In the model I have described, the
intermediate scale
symmetry breaking occurs even though no scalar squared
mass runs negative, and without tuning the relative magnitudes of any
superpotential terms. 
As is the case for the model in the previous section,
matter parity is automatically a conserved symmetry because of the way
$B-L$ is broken. In addition, the strong CP problem is solved and there is
an invisible axion (which is mainly
Im$[N- \overline N]/\sqrt{2}$) along the lines of ref.~\cite{kimnilles}.
The fermionic axino and scalar saxino partners of the axion both obtain
electroweak-scale masses, but they are only very weakly coupled to
ordinary matter. As before, one expects that the pattern of MSSM squark and
slepton masses will be augmented by a $B-L$ $D$-term contribution, which
could be distinguished at the CERN Large Hadron Collider or a future
lepton linear collider. As emphasized in section \ref{sec:toy}, the
magnitude and even the sign of the $D$-term depend on non-renormalizable
Kahler potential terms, but its existence is non-negotiable.

\section{Conclusions}
\label{sec:conclusions}
\setcounter{equation}{0}
\setcounter{footnote}{1}  

In this paper, I have studied the effects of supersymmetry-breaking
$\phi^4$ couplings in producing intermediate-scale VEVs. Any truly complete
model
of supersymmetry breaking should be able to predict the magnitude of
these terms, at least relative to the corresponding superpotential
couplings. As shown in sections 2 and 3, these 
terms are most important in or near $D$-flat directions that are also
$F$-flat at the renormalizable level. For these purposes, there is a
well-defined procedure for including RG effects, as given by the beta
function eq.~(\ref{rghgen}), which incorporates the large logarithms in
an effective potential approach.

An intermediate scale roughly of order $\sqrt{m_W M_{\rm
Planck}}$ is suggested both by the see-saw scenario for neutrino masses
and the allowed axion window \cite{axionwindow}. In sections 4 and 5,
I discussed simple models in which the existence of these scales is
dependent on the presence of dimensionless supersymmetry-breaking terms.
The numerical studies in this paper show that in the simplest cases,
one requires the values of the dimensionless ratio $h/xm_{1/2}$ at the
Planck scale
to be fairly large, and perhaps to favor a particular sign, in order
to achieve intermediate-scale symmetry breaking. However, this may well be
too conservative since Figs.~\ref{testvacgraph} and \ref{testvacnsgraph}
neglect the possible effects of negative contributions to scalar squared
masses from RG effects, even if the scalar squared masses never
run negative.
Furthermore, as I argued briefly in section 5, the parameter space in which
intermediate-scale symmetry breaking can take place can be increased
substantially if one is close to an accidental nearly-flat direction
and the $h \phi^4$ terms are 
misaligned with the corresponding superpotential terms. In 
any case, it is likely that the dimensionless supersymmetry
breaking couplings play a crucial role in producing VEVs along
supersymmetric flat directions.

\smallskip \noindent {\it Acknowledgements: } I thank Nir Polonsky, Pierre
Ramond and James Wells for useful comments, and the Aspen Center for
Physics for hospitality. This work was supported in part by National
Science Foundation grant number PHY-9970691.

\section*{Appendix: flat directions in the MSSM}
\label{sec:appendix}
\renewcommand{\theequation}{A.\arabic{equation}} 

In this paper, I have mainly been concerned with the ability of
dimensionless supersymmetry-breaking couplings to produce
desired intermediate-scale
VEVs like the PQ scale and the neutrino seesaw scale.
It is also interesting to consider the opposite case, when we do
not want an intermediate scale VEV to occur; namely, $h$-terms corresponding
to flat directions in the MSSM.
Supersymmetric flat
directions are
parameterized by analytic gauge-invariant polynomials in chiral
superfields.  For example, there are flat directions $QQQL$ and
$\overline u\overline u\overline d\overline e$
which are not lifted by any renormalizable supersymmetric terms in the
scalar potential. Instead, they are lifted 
by supersymmetry-breaking (mass)$^2$ terms, 
by non-renormalizable superpotential terms, 
and by the corresponding $h$-type terms. The latter terms always
lower the scalar potential for some choice of phases of the fields, and
therefore favor the existence of a non-trivial minimum at an intermediate
scale. Naively, these
minima are potentially dangerous, since they of course break color and
electric charge. However,
the mere existence of such vacua does not
imply that the universe has to be in one of them in the present day, even
if they are global minima. On the other hand, the dynamics associated with
the flat
directions may play an important role in the early universe, particularly
during inflation when the relevant $F$-term may be the auxiliary component
of the inflaton, and in the study of baryogenesis
\cite{Affleck:1985fy,Dine:1996kz}.
Here, I will illustrate the importance of the RG running by considering the
$h$-terms
for the $\overline u\overline u\overline d\overline e$ flat directions. The
analysis is very similar e.g.~for the $QQQL$ flat directions.

The relevant superpotential is given by
\beq
W = {\xx^{ijkl} \over 2 M} \overline u_i\overline u_j\overline d_k
\overline e_i
\eeq
where gauge indices are suppressed in the only possible way and $i=1,2,3$ 
is a family index. The supersymmetry-breaking
potential is given by
\beq
V = {h^{ijkl} \over 2 M} \overline u_i\overline u_j\overline d_k\overline
e_i
+ {\rm c.c.} +
\sum_{i=1}^3 \left (
m^2_{u_i} |\overline u_i|^2 +
m^2_{d_i} |\overline d_i|^2 +
m^2_{e_i} |\overline e_i|^2 \right ) .
\eeq
Now, suppose we are interested in a particular flat direction 
with $SU(3)_C$ color components
\beq
\overline u_1 = \pmatrix{ \phi \cr 0 \cr 0};\qquad
\overline u_2 = \pmatrix{ 0 \cr \phi \cr 0};\qquad
\overline d_1 = \pmatrix{ 0 \cr 0 \cr \phi};\qquad
\overline e_1 = \phi .
\eeq
Here the indices $1$ and $2$ on $\overline u_1$, $\overline u_2$,
$\overline d_1$ and $\overline e_1$ represent
arbitrarily-labelled family components. The potential along the flat
direction is
given by
\beq
V = 
\sum m^2 |\phi|^2 
+ \left ( {h\over M} \phi^4 + {\rm c.c.} \right ) +
{4 \overline x^2\over M^2}
|\phi|^6 ,
\eeq
where
\beq
\sum m^2 &\equiv& m_{u_1}^2 + m_{u_2}^2 +m_{d_1}^2 +m_{e_1}^2;\\
h &\equiv& h^{1211};\\
\overline x^2 &\equiv& {1\over 4} \sum_{i=1}^3 \left (
|x^{i211}|^2 + |x^{1i11}|^2 + |x^{12i1}|^2 + |x^{121i}|^2 \right ).
\eeq

In the MSSM, there are no Yukawa couplings which simultaneously link
any pair of fields $\overline u$, $\overline d$ and $\overline e$.
Furthermore, to a very good approximation, the Yukawa contribution to
wavefunction renormalizations are only non-vanishing for the
$t,b,\tau$ family. Therefore, the conditions leading to
eq.~(\ref{hoverxrunning})
are satisfied, and one can write a simple RG equation for each separate
ratio $h^{ijkl}/x^{ijkl}$:
\beq
{d\over dt} (h/x) = {g_3^2 \over \pi^2} M_3 + {3 g_1^2 \over 10 \pi^2}
M_1 .
\eeq
(I emphasize that this is true even when the coupling involves $t,b,\tau$
fields.)
The resulting running of this family-independent ratio is
graphed in Fig.~\ref{uudegraph}, starting from unified gauge couplings
at $Q_0 = 3 \times 10^{16}$ GeV. As
the graph shows, the ratio $(h/xm_{1/2})^2$
can become quite large for negative $h_0/x_0m_{1/2}$, while for positive
$h_0/x_0m_{1/2}$ it tends to run to very small values.
\begin{figure}[t]
\centering
\epsfxsize=4.1in
\hspace*{0in}
\epsffile{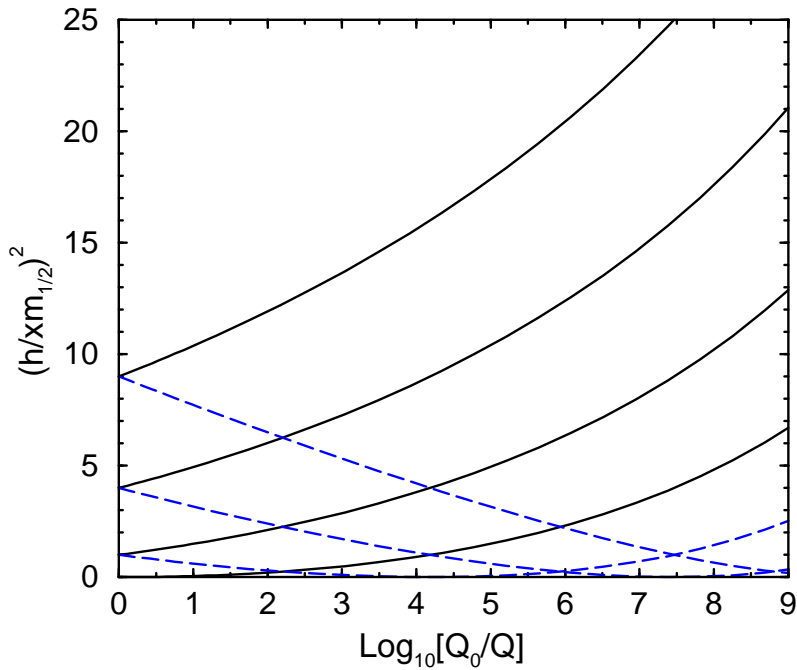}
\caption{RG running of $h/\xx m_{1/2}$ for a $
\overline u\overline u\overline d\overline e$ flat direction
in the MSSM as described in the text. 
Initial values 
at the high scale (corresponding to the left-hand side of the graph)
are $h/\xx m_{1/2} = 0,  -1, -2,  -3$ (solid lines) and
$h/\xx m_{1/2} = 1,2,3$ (dashed lines).
}
\label{uudegraph}
\end{figure}

A non-trivial local minimum in the scalar potential can occur at an
intermediate
scale if 
\beq
|h|^2 - 3 \overline x^2 \sum m^2 > 0.
\eeq
Note that while the RG running shown involves only
the coupling $x$ corresponding to the gauge-invariant polynomial of the flat
direction, the minimization condition that needs to be satisfied involves
the larger averaged squared coupling $\overline x^2$. If only one coupling
dominates, then $\overline x^2 \approx x^2$.
The relationship between $x^2$ and $\overline x^2$ is quite model-dependent;
however, there are strong constraints
on
these couplings because they are dimension 5 proton decay operators.
Therefore, it is probable that the largest of the couplings
$x$ belong to the $t,b,\tau$ family, which are constrained the least by
proton decay searches. In addition, the top and bottom squarks get
negative RG corrections to their masses due to the top and bottom Yukawa
couplings. Therefore, it is quite likely that if a non-trivial local minimum
exists, it will involve the third-family quarks and sleptons. I find that
such a minimum can easily exist for $h/xm_{1/2} \lsim -2$ or so,
because of the fast growth of $h/x$ in the infrared.
However, the precise regions of parameter space in which this can
occur depend on numerous unknown variables (including the scalar cubic
couplings) so I will not attempt to outline them here.


\end{document}